\documentclass[twocolumn,prb,showpacs,amsmath,amssymb]{revtex4-1}
\usepackage[dvipdfmx]{graphicx}
\usepackage{bm}
\begin{document}
\title{Counterflow quantum turbulence of He II in a square channel: Numerical analysis with nonuniform flows of the normal fluid}
\author{Satoshi Yui$^1$}
\author{Makoto Tsubota$^{1,2}$}
\affiliation{$^1$Department of Physics, Osaka City University, 3-3-138 Sugimoto, Sumiyoshi-Ku, Osaka 558-8585, Japan}
\affiliation{$^2$The OCU Advanced Research Institute for Natural Science and Technology (OCARINA), Osaka City University, 3-3-138 Sugimoto, Sumiyoshi-Ku, Osaka 558-8585, Japan}
\date{\today}

%**************************************************************************************************
%**************************************************************************************************
\begin{abstract}
  We perform a numerical analysis of counterflow quantum turbulence of superfluid $^4$He with nonuniform flows by using the vortex filament model.
  In recent visualization experiments  nonuniform laminar flows of the normal fluid, namely, Hagen-Poiseuille flow and tail-flattened flow, have been observed.
  Tail-flattened flow is a novel laminar flow in which the outer part of the Hagen-Poiseuille flow becomes flat.
  In our simulation, the velocity field of the normal fluid is prescribed to be  two nonuniform profiles.
  This work addresses a square channel to obtain important physics not revealed in the preceding numerical works.
  In the studies of the two profiles we analyze the statistics of the physical quantities.
  Under  Hagen-Poiseuille flow, inhomogeneous quantum turbulence appears as a statistically steady state.
  The vortex tangle shows a characteristic space-time oscillation.
  Under tail-flattened flow, the nature of the quantum turbulence depends strongly on that flatness.
  Vortex line density increases significantly as the profile becomes flatter, being saturated above a certain flatness.
  The inhomogeneity is significantly reduced  in comparison to the case of  Hagen-Poiseuille flow.
  Investigating the behavior of quantized vortices reveals that tail-flattened flow is an intermediate state between  Hagen-Poiseuille flow and uniform flow.
  In both profiles we obtain a characteristic inhomogeneity in the physical quantities, which suggests that a boundary layer of  superfluid appears near a solid boundary.
  The vortex tangle produces a velocity field opposite to the applied superfluid flow, and, consequently, the superfluid flow becomes smaller than the applied one.
\end{abstract}

\maketitle

%**************************************************************************************************
%**************************************************************************************************
\section{Introduction \label{introduction}}
Quantum turbulence in counterflow of He II has left a mystery:
 the transition between two different kinds of superfluid turbulence, namely, the T1-T2 transition \cite{tough}.
Recent developments in visualization techniques have enabled us to study this mystery.
However, our theoretical understanding has not yet been developed enough for two chief reasons:
1. the difficulty in taking proper account of the inhomogeneity of the system and the geometry of the channel in the formulation and
2. the need to solve the coupled dynamics of the two fluids to understand the phenomena fully.
In this paper, in order to address the first reason we perform a numerical analysis of the vortex dynamics by prescribing  realistic nonuniform profiles to the flow of the normal fluid.
The present work reveals characteristic inhomogeneous structure and dynamics of the vortex tangle that  has not been reported in preceding works with a uniform flow profile \cite{schwarz88} $^{,}$ \cite{adachi}.
We do not address directly the T1-T2 transition.
If we can address the coupled dynamics of the two fluids, we will be able to reveal the T1-T2 transition, and we are going to report on this work in the near future.

This study is motivated by the recent observation by Marakov {\it et al.} \cite{marakov} of the flow profiles of the normal fluid in thermal counterflow in a square channel.
They have observed the flow profiles of the normal fluid by following the motion of seeded metastable ${\rm He}_2^{*}$ molecules by a laser-induced-fluoresence technique \cite{guo09}.
Then in a thermal counterflow they prepare thin ${\rm He}_2^{*}$ molecular tracer lines perpendicular to the flow created by femtosecond-laser field ionization.
Above 1 K the molecules interact strongly with the normal fluid and trace the motion.
Thus how the tracer lines move reveals the flow profile of the normal fluid.
First, they presented evidence that the flow of the normal fluid is indeed turbulent at relatively large velocities \cite{guo10}.
In their recent studies they observe the novel transition of the flow profile of the normal fluid \cite{marakov}.
At relatively small velocities, the flow of the normal fluid remains laminar, whereas the superfluid is already turbulent.
For heat current $q < q_{c1}$, where the value of $q_{c1}$ obtained from the experiment is  $\sim$$50 ~\mathrm{mW/cm^2}$, an initially straight tracer line deforms to a nearly parabolic shape, indicating a laminar Poiseuille velocity profile.
For $q_{c1} < q < q_{c2}$, where the value of $q_{c2}$ obtained from the experiment is $80 ~\mathrm{mW/cm^2}$, the outer part of the tracer line becomes flattened, which is called the tail-flattened velocity profile.
Such a laminar profile has never been reported in classical fluid dynamics, and it must be characteristic of He II, namely, a result of the interaction between  the normal fluid and the superfluid.
In this work we study a numerical simulation of the vortex dynamics under these laminar profiles of the normal fluid.

Liquid $^4$He shows superfluidity below a $\lambda$ point of $2.17 ~{\rm K}$.
In a two-fluid model one considers superfluid $^4$He as an intimate mixture of  normal fluid and  superfluid components:
the normal fluid has a viscosity, whereas the superfluid is  inviscid.
The velocities of the normal fluid and the superfluid component are denoted by ${\bm v}_n$ and ${\bm v}_s$, respectively, and the densities are denoted by $\rho_n$ and $\rho_s$, respectively.
The total density $\rho=\rho_n + \rho_s$ is almost independent of temperature.
The ratio of the two fluids depends on temperature, and $\rho_s/\rho$ increases as temperature decreases.
The velocity fields obey different two equations of motion \cite{donnelly}, and they are basically decoupled.

The idea of quantized circulation in superfluid helium was suggested by Onsager \cite{onsager} and has been observed experimentally by Vinen \cite{vinen61}.
In a superfluid any rotational motion is sustained only by quantized vortices, which have the quantum circulation $\kappa = h/m_4$, where $h$ is Planck's constant and $m_4$ is the mass of a $^4 {\rm He}$ atom.
The elementary excitations forming the normal fluid are strongly scattered by these quantized vortices.
Thus, if there is a relative velocity between the normal fluid and the quantized vortices, a frictional force works between them, being called a mutual friction force.
In the superfluid turbulent state the mutual friction term is added to the equations of motion of the two fluids, so that the two fluids become coupled.

Thermal counterflow in He II, which is  internal convection of the two fluids, has been studied well as a prototype of quantum turbulence.
One end of the channel is connected with a He II bath and the other end is closed.
Upon heating the closed end, the normal fluid flows to the cooler side to transfer  heat.
The superfluid flows oppositely to the warmer end to satisfy  mass conservation:
\begin{equation}
  \int ( \rho_n {\bm v}_n + \rho_s {\bm v}_s ) dS = {\bm 0} ,
  \label{mass}
\end{equation}
where the integral is performed over the cross section of the channel.
Thus a relative velocity $v_{ns} = \overline{ |{\bm v}_{n}-{\bm v}_{s}| }$ occurs between the two fluids, where the overline denotes the spatial average over the channel cross section.
When the counterflow velocity exceeds a critical value, a self-sustaining tangle of quantized vortices appears to form superfluid turbulence.
The measurements \cite{vinen57} show that the vortex line density $L$ follows the relation
\begin{equation}
  L^{\frac{1}{2}} = \gamma (v_{ns} - v_0),
  \label{vinen}
\end{equation}
where $\gamma$ is a parameter depending on temperature and $v_0$ is a small parameter.

The scheme for understanding quantum turbulence from the vortex dynamics was addressed by Vinen.
By assuming homogeneous turbulence he estimated the vortex growth by using a dimensional analysis and modeled the decay process phenomenologically \cite{vinen57}.
He showed that the dynamics of the vortex tangle is described by  Vinen's equation
\begin{equation}
  \frac{dL}{dt} = \chi_1 v_{ns} L^{\frac{3}{2}} - \chi_2 L^{2} ,
\end{equation}
where $\chi_1$ and $\chi_2$ are temperature-dependent parameters.
If the vortex tangle is in a steady state, $dL/dt =0$, resulting in Eq. (\ref{vinen}).

Schwarz has investigated counterflow quantum turbulence using the vortex filament model \cite{schwarz88}.
The observable quantities obtained in his calculation agree with the experimental results for the steady state of vortex tangles.
However, his simulation could sustain the steady vortex tangle only through some artificial procedure.
Adachi {\it et al.} argued that this comes from the localized induction approximation (LIA) in which the interaction between vortices is neglected \cite{adachi}.
By performing the full Biot-Savart calculation they removed the difficulty and successfully obtained the steady state consistent with the experimental results.
We note that in these numerical works  the system was assumed to be uniform and the physics coming from the inhomogeneity was missed; i.e., the effects of the channel walls and the resulting flow profile were missed.

Counterflow quantum turbulence is known to depend on the aspect ratio of the cross section of the channel \cite{tough}.
In low-aspect-ratio channels, the system has  two turbulent states.
An increase in the counterflow velocity is observed to change the laminar state to the first turbulent state T1, and next to the second turbulent state T2.
Melotte and Barenghi \cite{melotte} suggested that the transition from  T1 to T2 is caused by the transition of the normal fluid from laminar to turbulent.
They addressed the issue of the stability of the normal fluid with the relation to the T1-T2 transition.
In high-aspect-ratio channels, the counterflow exhibits only a single turbulent state T3, but there has been little information about what  T3 is.

The recent visualization experiments have made important breakthroughs for these issues.
By using  micron-sized solid hydrogen tracers the Maryland group \cite{bewley} succeeded in visualizing  quantized vortices in superfluid $^4$He.
Then they applied the technique to thermal counterflow to confirm directly the behavior of the two-fluid model \cite{paoletti}.
Zhang {\it et al.} have studied a counterflow channel containing a cylinder and showed that macroscopic eddies appear both downstream and upstream of the cylinder \cite{zhang}.
La Mantia {\it et al.} have investigated the Lagrangian dynamics of micrometer-sized solid particles in thermal counterflow and obtained the normalized probability functions of the particle velocities and accelerations \cite{mantia}.
Recently,  visualization of  normal fluid flow in  thermal counterflow has been performed.
Marakov {\it et al.} \cite{marakov} found that the normal fluid in a thermal counterflow has  two characteristic laminar flows, as described in the second paragraph of this introduction.
Recent developments in cryogenic visualization techniques have shed light on the problems of inhomogeneous quantum turbulence.

We mention  previous numerical studies of  thermal counterflow in He II with a nonuniform profile.
Samuels \cite{samuels} studied the dynamics of quantized vortices in laminar circular pipe flow with  normal fluid Poiseuille flow and superfluid flat flow in the same direction.
Aarts {\it et al.} \cite{aarts} studied the vortex tangle in  thermal counterflow by assuming a normal-component velocity profile ether flat or parabolic in variously shaped channels with the LIA.
Baggaley {\it et al.} \cite{baggaley13}$^{,}$\cite{baggaley14} studied  thermal counterflow between two parallel plates by using the prescribed Poiseuille and turbulent profiles for the normal fluid flow.

A vortex tangle obtained by Baggaley {\it et al.} \cite{baggaley13} concentrated near the channel walls and showed characteristic inhomogeneity.
The simulation generated  quantum turbulence under  laminar Poiseuille flow and turbulent flow of the normal fluid.
By examining results, they argued that their results supported the scenario proposed by Melotte and Barenghi.
A better understanding of T1 and T2 states would be obtained by studying the flow in a low-aspect-ratio channel where all boundaries are solid except for the flow direction.
This is because the T1 and T2 states are actually observed in low-aspect-ratio channels and another turbulent state T3 is observed in high-aspect-ratio channels \cite{tough}.

In this paper we performed a numerical analysis of  thermal counterflow in a square channel.
It is necessary to address the coupled dynamics of the two fluids to understand the T1-T2 transition.
In a typical model of the coupled dynamics one would suppose that the superfluid is described by the vortex filament model, the normal fluid is described by the Navier-Stokes equation, and they are coupled through  mutual friction \cite{kivotides}.
However, it is difficult to solve fully the coupled dynamics.
As the first essential step, we address the T1 state where the normal fluid must be laminar.
The preliminary results for Hagen-Poiseuille flow was reported in the previous study \cite{yui}.
The normal fluid is prescribed to be two types of realistic flow: Hagen-Poiseuille flow and tail-flattened flow.
In both  profiles, an inhomogeneous vortex tangle is obtained as a statistically steady state, where a boundary layer of the superfluid appears near the channel walls.
Under Hagen-Poiseuille flow the vortex tangle shows a characteristic space-time oscillation.
Tail-flattened flow was just observed by Marakov {\it et al.} \cite{marakov}, and its nature is unknown.
The present simulation reveals that tail-flattened flow is some intermediate state between Hagen-Poiseuille flow and uniform flow.

The contents of this paper are as follows.
Section II describes the formulation of this work.
In Sec. III we show results of our  simulation with  Hagen-Poiseuille flow.
In Sec. IV we show results of our simulation with tail-flattened flow.
Section V is devoted to a discussion.
Section VI presents our conclusions.

%**************************************************************************************************
%**************************************************************************************************
\section{Formulation \label{formulation}}
Here we describe the formulation of a vortex filament model and its numerical analysis.
The equation of motion of a vortex filament is first explained,
then the velocity fields of the normal fluid are described and the
conditions of our simulation are noted.
The physical quantities necessary for characterizing the vortex tangle are also introduced.

%--------------------------------------------------------------------------------------------------
\subsection{Equation of motion}
In a vortex filament model \cite{schwarz85} a quantized vortex is represented by a filament passing through a fluid and has a definite vorticity.
This approximation is very suitable in He II, because the core size of a quantized vortex is much smaller than any other characteristic length scale.

At zero temperature the vortex filament moves with the superfluid velocity
\begin{equation} \label{super}
  {\bm v}_s =
      {\bm v}_{s,\omega}
    + {\bm v}_{s,b}
    + {\bm v}_{s,a}
    ,
\end{equation}
where ${\bm v}_{s,\omega}$ is the velocity field produced by vortex filaments, ${\bm v}_{s,b}$ is that produced by solid boundaries, and ${\bm v}_{s,a}$ is the applied uniform flow of the superfluid.
The velocity field ${\bm v}_{s,\omega}$ is given by the Biot-Savart law
\begin{equation}
  {\bm v} _{s,\omega} ({\bm r})=
    \frac{\kappa}{4 \pi} \int _{\cal L}
      \frac{ ({\bm s} _{1} - {\bm r}) \times d {\bm s} _{1} }{ |{\bm s} _{1} - {\bm r} |^{3} }
      .
\end{equation}
The filament is represented in parametric form as ${\bm s}={\bm s}(\xi,t)$, where ${\bm s}_1$ refers to a point on the filament and the integration is performed along the filament.
This work addresses the full Biot-Savart integral \cite{adachi}.
The velocity field ${\bm v}_{s,b}$ is obtained by a simple procedure;
it is just the field produced by an image vortex that is constructed by reflecting the filament into the surface and reversing its direction.
According to Eq. (\ref{mass}) we set $v_{s,a} = \rho_{ns} \overline{v_n}$ in thermal counterflow, where $\rho_{ns} = \rho_n / \rho_s$.

By taking into account mutual friction, the velocity of a point $\bm s$ on the filament is given by
\begin{equation}
  {\dot{\bm s}} = {{\bm v}_s}
    + \alpha {\bm s}' \times ({\bm v}_n - {{\bm v}}_s)
    - \alpha '{\bm s}' \times [{\bm s}'\times({\bm v}_n - {{\bm v}}_s)],
  \label{vortex}
\end{equation}
where $\alpha$ and $\alpha'$ are the temperature-dependent coefficients, and the prime denotes derivatives of ${\bm s}$ with respect to the coordinate $\xi$ along the filament.

%--------------------------------------------------------------------------------------------------
\subsection{Flow profiles of the normal fluid}
The velocity ${\bm v}_n$ of the normal fluid is prescribed to be  two profiles, namely, the Hagen-Poiseuille profile $u_{p}$ and the tail-flattened $u_{t}$ profile.
We suppose that the flow direction is along the $x$ coordinate and that $y,z$ are coordinates are normal to $x$, namely, ${\bm v}_n = u(y,z) {\hat{\bm x}}$;  the center of the channel is $(y,z)=(0,0)$.

In a rectangular channel the Hagen-Poiseuille profile is given by
\begin{equation}
\begin{split}
  u_{p}(y,z) =& u_0 \sum _{m=1,3,5,\cdots} ^{\infty} (-1)^{\frac{m-1}{2}}
  \\
  &
  \times
  \left[
    1 - \frac{ \cosh(m \pi z / 2 a) }{ \cosh(m \pi b / 2 a) }
  \right]
  \frac{ \cos(m \pi y / 2 a) }{ m^3 }
  ,
\end{split}
\label{poi}
\end{equation}
where $u_0$ is a normalization factor and $a \text{ and }b$ are halves of the channel width along the $y \text{ and }z$ axes, respectively \cite{poiseuille}.

\begin{figure}[ht]
  \includegraphics[width=0.35\textwidth]{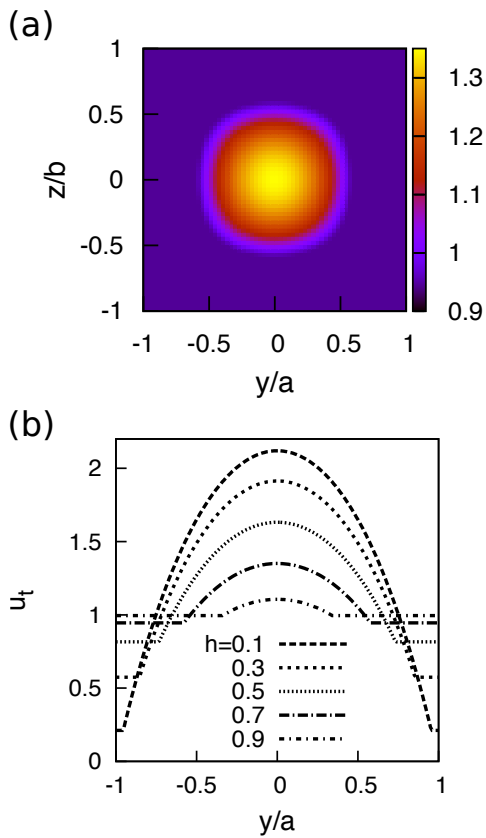}
  \caption
  {
    (Color online)
    Tail-flattened profile $u_t$, normalized by the mean value.
    (a)
    Profile with $h=0.7$.
    The color bar shows the value of $u_t$.
    (b)
    Profiles along $z=0 ~\mathrm{mm}$ with various values of $h$.
  }
  \label{fla_profile}
\end{figure}

We realize the tail-flattened profile by composing the expression\begin{equation}
  u_{t}(y,z) = u_0 \mathrm{max}[ u_{p}(y,z), h u_{p}(0,0) ]
  ,
\label{fla}
\end{equation}
where $\mathrm{max}[A,B]$ refers to the larger value of the two arguments, and $0 < h < 1$ is a fitting parameter, which determines the flattened area.
Figure \ref{fla_profile}(a) shows the profile of tail-flattened flow over the channel cross section, and Fig. \ref{fla_profile}(b) shows the profiles along $z=0 ~\mathrm{mm}$ with various $h$ values.
$u_t$ is equal to $u_p$ for $h=0$, and $u_t$ becomes  uniform for $h=1,$ corresponding to simulation by Adachi ${\it et~al.}$ \cite{adachi}.
The increase in $h$ makes $u_t$ more uniform, as shown in Fig. \ref{fla_profile}(b).

%--------------------------------------------------------------------------------------------------
\subsection{Conditions}
The simulations are performed under the following conditions.
We discretize the vortex lines into a number of points held at a minimum space resolution of $\Delta \xi = 8.0 \times 10^{-4} ~\mathrm{cm}$.
Integration in time is achieved using a fourth-order Runge-Kutta scheme with  time resolution $\Delta t = 1.0 \times 10^{-4} ~\mathrm{s}$.
The computing box is $0.1 \times 0.1 \times 0.1 \mathrm{~cm^3}$.
Periodic boundary conditions are used along the flow direction $x$, whereas solid boundary conditions are applied to the channel walls at $y=\pm 0.05 ~\mathrm{cm}$ and $z=\pm 0.05 ~\mathrm{cm}$.
The solid boundaries are assumed to be smooth completely, thus neither vortex pinning nor surface drag occurs.
We reconnect two vortices artificially, when the vortices approach each other more closely than $\Delta \xi$.
We eliminate vortices that are shorter than $5 \times \Delta \xi = 2.4 \times 10^{-3} ~\mathrm{cm}$.
Temperatures \cite{schwarz85} are $T=1.9 ~\mathrm{K} ~(\alpha=0.21, \alpha'=0.009)$, $1.6 ~\mathrm{K} ~(\alpha=0.098, \alpha'=0.016)$, and $1.3 ~\mathrm{K} ~(\alpha=0.036, \alpha'=0.014)$.
The density of the normal fluid \cite{barenghi} is $\rho_n/\rho= 0.43$ at $1.9 ~\mathrm{K}$, $0.16$ at $1.6 ~\mathrm{K}$, $0.05$ at $1.3 ~\mathrm{K}$.
The initial state consists of eight randomly oriented vortex rings of radius = $0.023 \mathrm{~cm}$.

%--------------------------------------------------------------------------------------------------
\subsection{Physical quantities \label{physical}}
In this subsection we introduce the physical quantities used in this paper.

The vortex line density is defined as
\begin{equation}
  L = \frac{1}{\Omega} \int _{\cal L} d\xi,
\end{equation}
where the integral is performed along all vortices in the sample volume $\Omega$.

The anisotropy of the vortex tangle is represented by  dimensionless parameter \cite{schwarz88}:
\begin{equation}
  I = \frac{1}{\Omega L} \int _{\cal L} [1-({\bm s}' \cdot \hat{\bm r}_{\mathrm p})^2 ] d\xi.
\end{equation}
Here $\hat{\bm r}_{\mathrm p}$ represents the unit vector parallel to the flow direction.
If the vortex tangle is completely isotropic, $I = 2/3$.
If the tangle consists entirely of curves lying in the plane normal to the flow direction, $I=1$.

To analyze the inhomogeneity of the vortex tangle, we divide the computational box into subvolumes by using a uniform $31^3$ Cartesian mesh.
The vortex line density is defined at a subvolume as the local vortex line density
\begin{equation}
  L' = \frac{1}{\Omega'} \int _{\cal L'} d\xi,
\end{equation}
where the integral is performed along all vortices in the subvolume $\Omega '$.
The local anisotropic parameters  are defined as
\begin{equation}
  I' = \frac{1}{\Omega' L'} \int _{\cal L'} [1-({\bm s}' \cdot \hat{\bm r}_{\mathrm p})^2 ] d\xi.
\end{equation}
These local quantities are always averaged over the flow direction $x$.
Thus these are functions of position $y, z$ normal to the flow direction, namely, $L'(y,z)$ and $I'(y,z)$.

%**************************************************************************************************
%**************************************************************************************************
\section{Hagen-Poiseuille flow \label{poiseuille}}
In this section we present results of our simulation under the Hagen-Poiseuille flow expressed by Eq. (\ref{poi}).
We analyze physical quantities to discuss the dynamical and statistical properties of the quantum turbulence.

%--------------------------------------------------------------------------------------------------
\subsection{Dynamics}
We discuss the time development of $L$, and we show that the inhomogeneous vortex tangle is sustained by space-time oscillation.
This is characteristic of the present case, which cannot appear in  uniform counterflow \cite{adachi}.

\begin{figure}[b]
  \includegraphics[width=0.4\textwidth]{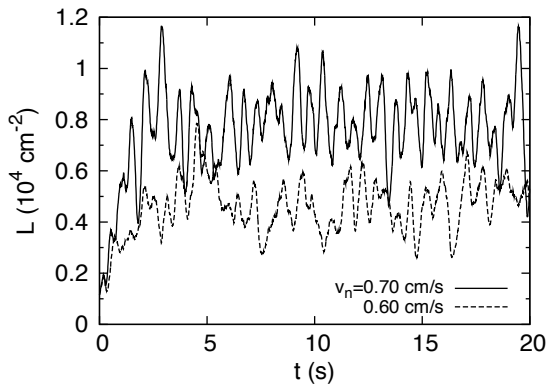}
  \caption
  {
    Vortex line density as a function of $t$ under  Hagen-Poiseuille flow with $T=1.9 ~\mathrm{K}$.
  }
  \label{poi_vld_19k.eps}
\end{figure}

Figure \ref{poi_vld_19k.eps} shows the time evolution of $L$.
The temperature is $T=1.9 ~\mathrm{K}$, and the mean velocity of the normal fluid is $v_n=0.6 \mathrm{}$ and $0.7 ~\mathrm{cm/s}$.
The vortex tangle develops to the statistically steady state, even if the counterflow is nonuniform.
Fluctuations in $L$ are much larger than those in  uniform counterflow \cite{adachi}.
The amplitude of the fluctuation increases with $v_n$.
The large fluctuation would imply that this system is unstable.
To investigate the origin of the large fluctuation, we focus on the configuration of the vortex tangle.
The large fluctuation of $L$ is attributable to a space-time oscillation coming from the inhomogeneous dynamics of the vortex tangle.
In the following the details of the oscillation are explained.

\begin{figure}[ht]
  \includegraphics[width=0.45\textwidth]{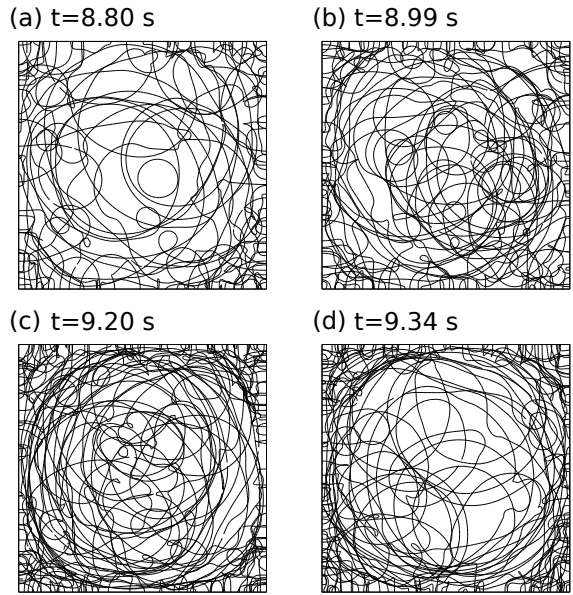}
  \caption
  {
  Vortex tangles viewed along the flow direction in the statistically steady state for $T=1.9~\text{K}$ and $v_{n}=0.7~{\rm cm/s}$, corresponding to the results of Fig. \ref{poi_vld_19k.eps}.
  These show one cycle of the large oscillation, while snapshots from (a) to (d) correspond a local minimum ($t=8.80 ~\mathrm{s}$), the middle of the increase ($t=8.99 ~\mathrm{s}$ ), a local maximum ($t=9.20 ~\mathrm{s}$), and the middle of the decrease ($t=9.34 ~\mathrm{s}$), respectively.
  }
  \label{tangles}
\end{figure}

Figure \ref{tangles} shows the space-time pattern of the vortex tangle at $T= 1.9 ~\mathrm{K}$ and $v_n=0.7 ~\mathrm{cm/s}$.
The period of the space-time oscillation is about $0.7 ~\mathrm{s}$, and the oscillation consists of four stages (a)--(d).
In Fig. \ref{tangles}(a), corresponding to a local minimum of $L$, the vortices are dilute, and the vortices remain only near the solid walls.
Then the vortices near the walls invade  the central region in Fig. \ref{tangles}(b).
These  vortices make numerous reconnections in the central region subject to the large counterflow in Fig. \ref{tangles}(c).
Hence $L$ increases significantly to a local maximum.
Eventually, in Fig. \ref{tangles}(d), the Hagen-Poiseuille flow excludes the vortex tangle from the central region toward the solid walls.
Thus the vortices in the central region become dilute, and they are absorbed by the solid boundaries.
Even if the vortex tangles of (b) and (d) have the almost same values of $L$, the vortex configurations are quite different.
Then the vortex tangle returns to the stage of Fig. \ref{tangles}(a).
The vortex tangle repeats the periodic motion, resulting in the large oscillation of Fig. \ref{poi_vld_19k.eps}.
The vortex tangle sustained by this mechanism is more dilute than that in the case of uniform counterflow for the same $T$ and $v_{ns}$.

%--------------------------------------------------------------------------------------------------
\subsection{Statistics}
We investigate the statistics of the quantum turbulence in a statistically steady state.
In this subsection every physical quantity is averaged  temporally over the statistically steady states and spatially over the computational box.
Typical experiments using second sound attenuation have observed values integrated over the channel.
The statistical values have been extensively investigated both experimentally and numerically.
We discuss the properties of  Hagen-Poiseuille normal fluid flow by comparing the statistical values with those obtained in previous studies of uniform counterflow.

\begin{figure}[ht]
  \includegraphics[width=0.4\textwidth]{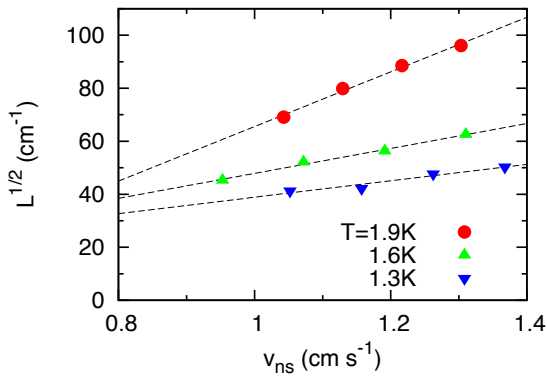}
  \caption
  {
    (Color online)
    Vortex line density averaged over the statistically steady state under  Hagen-Poiseuille flow as a function of $v_{ns}$.
    The dashed lines refer to the fitting of the data for each temperature.
  }
  \label{poi_vld_ave.eps}
\end{figure}

\begin{table}[ht]
  \caption{
    Line density coefficients $\gamma$, fitting parameter $v_0$ and anisotropic parameter $I$.
    Given are numerical results $\gamma_{\rm hp}$, $I_{\rm hp}$ and $v_0$ obtained in this work under Hagen-Poiseuille flow and $\gamma_{\rm uni}$ and $I_{\rm uni}$ obtained by Adachi {\it et al.} \cite{adachi}, which is a simulation subject to uniform counterflow under a periodic boundary condition.
  }
  \begin{tabular}{cccccc}
    \hline\hline
    $T$& $\gamma_{\rm hp}$ & $\gamma_{\rm uni}$ & $v_0$ & $I_{\rm hp}$ & $I_{\rm uni}$ \\
    (K)& (s/cm$^2$) &(s/cm$^2$)  &(cm/s) & & \\
    \hline
      1.3 & 31    & 53.5   & $-$0.2  & 0.765 & 0.738 \\
      1.6 & 47    & 109.6  &  0.0    & 0.814 & 0.771 \\
      1.9 & 103   & 140.1  &  0.4    & 0.878 & 0.820 \\
    \hline\hline
  \end{tabular}
  \label{gamma}
\end{table}

The statistically steady state is known to satisfy the relation of Eq. (\ref{vinen}).
Here we regard the counterflow velocity $v_{ns}$ as the spatially averaged amplitude of ${\bm v}_{ns}={\bm v}_{n}-{\bm v}_{s,a}$.
Figure \ref{poi_vld_ave.eps} shows that $L$ almost satisfies the relation.
The vortex line density increases with $T$, because the mutual friction becomes strong at higher $T$ to develop quantized vortices.
The critical velocity $v_c$ below which the vortex tangle cannot be sustained is not investigated in this work, since the attempt to determine $v_c$ correctly is time consuming and  is not the main purpose of this work.
Table \ref{gamma} shows a comparison of $\gamma$ between the present work, $\gamma_{\rm hp}$, and the simulation, $\gamma_{\rm uni}$, subject to  uniform counterflow under a periodic boundary condition \cite{adachi}.
The values of $\gamma_{\rm uni}$ quantitatively agree with those of the typical experiment \cite{childers}; thus we can regard $\gamma_{\rm uni}$ as a standard.
The values of $\gamma_{\rm hp}$ are lower than those of $\gamma_{\rm uni}$.
This suggests that the Hagen-Poiseuille flow tends to suppress the growth of the vortex tangle.
The fitting parameters $v_0$ of the present simulation are listed in Table \ref{gamma}.
These results are different from the value $v_0 \sim 0.1 ~\mathrm{cm/s}$ obtained by the simulation under  uniform periodic counterflow \cite{adachi}.

\begin{figure}[ht] 
  \includegraphics[width=0.4\textwidth]{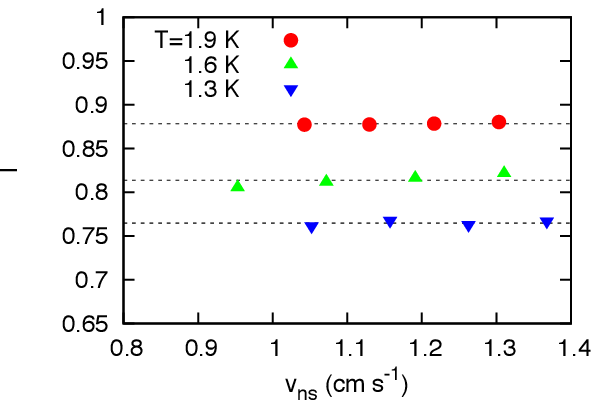} 
  \caption{ 
    (Color online)
    Anisotropic parameter $I$ averaged over the statistically steady state under Hagen-Poiseuille flow as a function of $v_{ns}$.
    The dotted lines are the values of $I$ averaged over $v_{ns}$ for each temperature.
  }
  \label{poi_iso_ave.eps}
\end{figure}

Figure \ref{poi_iso_ave.eps} shows the anisotropic parameter $I$ as a function of $v_{ns}$ and $T$.
The anisotropy is almost independent of $v_{ns}$, in agreement with experiments \cite{wang}.
The dotted lines are the values of $I$ averaged over $v_{ns}$ for each temperature.
The anisotropy becomes large with $T$, because the increase in $T$ intensifies the expansion of vortices toward the direction normal to the flow by  mutual friction.
In Table \ref{gamma} we compare our results with the mean values of $I$ obtained by Adachi {\it et al.} \cite{adachi}, which is a simulation subject to uniform counterflow in a periodic cube.
The results with Hagen-Poiseuille flow are higher than those with uniform flow.
Thus the Hagen-Poiseuille flow tends to increases the anisotropy.
The difference between the results of this work and those from uniform counterflow increases with $T$.
This is because  the mutual friction becomes stronger with $T$ to enhance the effect of the Hagen-Poiseuille profile.

%--------------------------------------------------------------------------------------------------
\subsection{Inhomogeneity}
We have discussed the physical quantities averaged over the whole volume.
Here we analyze local physical quantities introduced in Sec. \ref{physical} to investigate the structure of the vortex tangle.
In this subsection  every physical quantity is averaged temporally over the statistically steady states and spatially over the flow direction.

\begin{figure*}[ht]
  \includegraphics[width=0.9\textwidth]{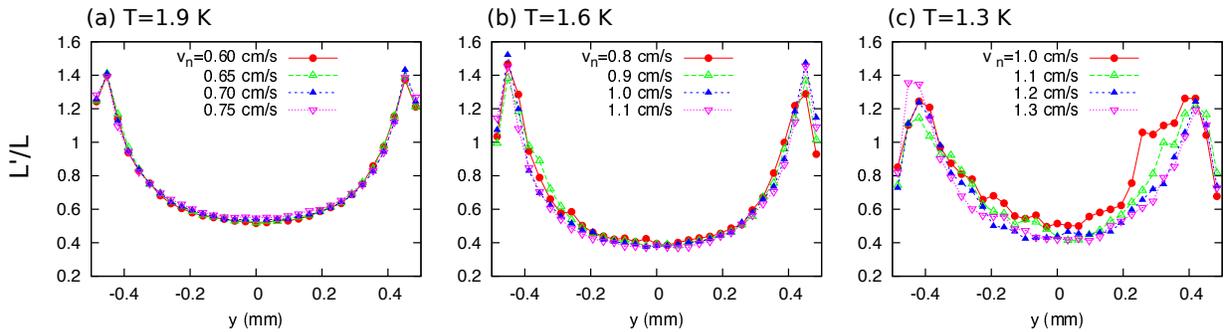}
  \caption
  {
    (Color online)
    Profiles of the vortex line density $L'$ normalized by $L$ along $z=0 ~\mathrm{mm}$ under Hagen-Poiseuille flow.
    Temperatures are $T=1.9$, $1.6 $, and $1.3 ~\mathrm{K}$ from left to right.
  }
  \label{vld3_poi}
\end{figure*}

\begin{figure*}[ht]
  \includegraphics[width=0.9\textwidth]{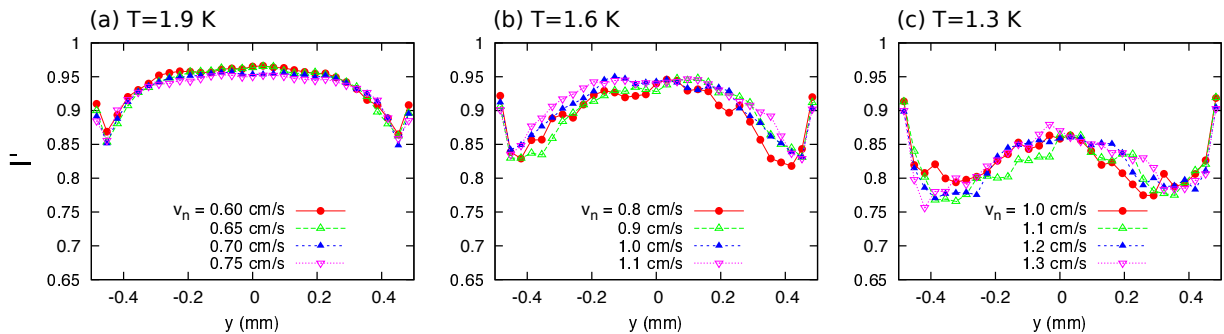}
  \caption
  {
    (Color online)
    Profiles of the anisotropic parameter $I'$ along $z=0 ~\mathrm{mm}$ under  Hagen-Poiseuille flow.
    Temperatures are $T=1.9$, $1.6 $, and $1.3 ~\mathrm{K}$ from left to right.
  }
  \label{iso3_poi}
\end{figure*}
Figure \ref{vld3_poi} shows the spatially dependence of the local vortex line density $L'$ normalized by $L$.
These are profiles along $z=0 ~\mathrm{mm}$ at three temperatures, $T=1.9$, $1.6$, and $1.3 ~\mathrm{K}$ (from  left to  right).
Within each subfigure, we plot the profiles for four different values of $v_{n}$, and the profile is almost independent of $v_{n}$.
One can see that the vortices concentrate near the channel walls.
The region with large values of $L'$ can be called a boundary layer of the superfluid.
Such a boundary layer of the superfluid was already found by Baggaley {\it et al.} in their simulation of counterflow between parallel plates \cite{baggaley13}$^{,}$\cite{baggaley14}.
This comes from analogy with the boundary layer of a viscous fluid, which is the layer near the surface where the effects of viscosity are significant and the intensity of vorticity is strong.
As shown in Sec. \ref{super}, indeed the vorticity of the superfluid increases in the boundary layer.
The vortex line density decreases very close the channel walls.
This is a consequence of the quantized vortices entering the surface normally to satisfy the solid boundary condition \cite{schwarz85}.
This effect tends to exclude vortices parallel to the surface, leading to the decrease in $L'$.

We discuss the dependence of the $L'$ profile on $T$ in terms of the difference between the maximum and minimum values of $L'$.
The nonuniform ${\bm v}_{ns}$ makes the vortex tangle inhomogeneous through the mutual friction terms of $\alpha$ and $\alpha'$ in Eq. (\ref{vortex}).
Both the counterflow velocity profile and the mutual friction coefficients depend on $T$.
First, the nonuniformity of ${\bm v}_{ns}$ decreases with $T$, reducing the inhomogeneity of the vortex tangle.
According to Eq. (\ref{mass}), the counterflow velocity can be written as ${\bm v}_{ns} = {\bm v}_n - {\bm v}_{s,a} = {\bm v}_{n} + \rho_{ns} \overline{\bm v}_{n}$, where $\rho_{ns} = \rho_n / \rho_s$.
The value of $\rho_{ns}$ increases with $T$, making ${\bm v}_{ns}$ more uniform, because the contribution of nonuniform ${\bm v}_{n}$ decreases.
Second, the coefficient $\alpha$ increases with $T$, enhancing the inhomogeneity of the vortex tangle.
The coefficient $\alpha'$ is a more complicated function of $T$.
These effects could maximize the difference between the maximum and minimum values of $L'$ at some temperature.
According to our results, the temperature may be close to $T=1.6 ~\mathrm{K}$.

Figure \ref{iso3_poi} shows the spatial dependence of the local anisotropic parameter $I'$.
These are profiles along $z=0 ~\mathrm{mm}$ at three temperatures, $T=1.9 $, $1.6 $, and $1.3 ~\mathrm{K}$ (from  left to  right).
Within each subfigure, we plot the profiles for four different values of $v_{n}$, and the profile is almost independent of $v_{n}$.
We can see that $I'$ values near the channel walls are lower than those in the central region.
This shows that the vortices in the boundary layer tend to be isotropic, compared with those in the central region.
With increasing $T$, the boundary layer becomes wider, which is also shown in Fig. \ref{vld3_poi}.
At the channel walls the anisotropy becomes significantly larger.
This is because the quantized vortices enter the surface normally to satisfy the solid boundary condition, as discussed above.

%--------------------------------------------------------------------------------------------------
\subsection{Superfluid velocity field \label{super}}
It is important to know how the superfluid velocity field ${\bm v}_s$ behaves in quantum turbulence.
This issue is closely related to the classification of two types of turbulence in He II at finite temperatures.
One type is that in which both the superfluid and the normal fluid are turbulent and couple strongly, where the coupled turbulence exhibits the Kolmogorov law for its energy spectrum \cite{maurer}$^{,}$\cite{stalp}.
The other is, like thermal counterflow, that in which the mean superfluid and the normal fluid velocities are not necessarily the same.
In any case, it is important to know how the superfluid follows the normal fluid through  mutual friction.
Samuels reported the velocity matching of the two fluids in laminar circular pipe flow with normal fluid Poiseuille flow and superfluid flat flow in the same direction \cite{samuels}.
In our system, the two fluids tend to mimic each other through  mutual friction, but a forced relative motion in the thermal counterflow prevents their velocity matching.
Then the nature of ${\bm v}_s$ is nontrivial.

\begin{figure*}[ht]
  \includegraphics[width=0.9\textwidth]{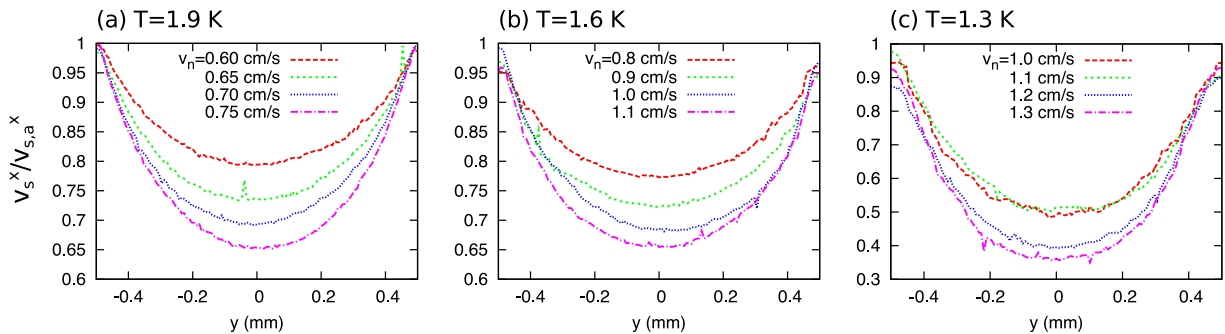}
  \caption
    {
    (Color online)
    Flow direction component $v_s^x$ of the velocity fields ${\bm v}_s$ of the superfluid along $z=0 ~\mathrm{mm}$ under  Hagen-Poiseuille flow.
    These values are normalized by $v_{s,a}^x$.
    Temperatures are $T=1.9$, $1.6$, and $1.3 ~\mathrm{K}$ from  left to  right.
    }
  \label{prt2_poi}
\end{figure*}

The flow direction component $v_s^x$ of the superfluid velocity field ${\bm v}_{s}$ is plotted along $z=0 ~\mathrm{mm}$ in Fig. \ref{prt2_poi} at temperatures of $T=1.9$, $1.6 $, and $1.3 ~\mathrm{K}$ (from left to  right).
These values are normalized by $v_{s,a}^x$.
Within each subfigure, we plot the profiles for four different velocities.
The outstanding feature through all these data is the velocity reduction in the central region.
The reduction is caused by the large velocity gradient near the channel walls, meaning that the intensity of vorticity near the channel walls is stronger.
This comes from the characteristic profiles of $L'$ and $I'$.
Also we can understand the velocity reduction in the central region by considering the Biot-Savart law.
Investigating in detail the vortex structure of Fig. \ref{tangles} finds that in the boundary layer most of vortices have some small radius of curvature comparable to the layer width.
These vortices propagate against $v_n$, and thus they create a velocity field parallel to $v_n$ in the central region.
Hence the quantized vortices in the boundary layer lead to the reduction in the central region.
The dependence of $v_s^x$ on $T$ appears mainly as the difference of the reduction integrated over the cross section.
This will be discussed in Sec. \ref{modification}.
Finally, we explain the dependence of $v_s^x$ on the value of $v_n$.
The velocity reduction of the superfluid becomes large with $v_n$.
Obviously, this comes from the increase in $L$.
As shown above, $L$ increases with $v_n$, even if the profile of $L'$ does not depend on $v_n$.
Thus the intensity of the vorticity in the boundary layer increases with $v_n$, so that the reduction of $v_s^x$ becomes larger.

The question we have to consider is whether or not the velocity reduction occurs in  uniform counterflow.
We conducted another simulation under  uniform counterflow in a periodic cube for $T=1.9 ~\mathrm{K}$ and $v_n=0.2 ~\mathrm{cm/s}$.
The obtained profile of $v_s^x$ is almost equal to that of $v_{s,a}$.
This shows that the reduction in the superfluid flow cannot be realized, unless there are a solid boundary and a resulting nonuniform flow of the normal fluid.

%**************************************************************************************************
%**************************************************************************************************
\section{tail-flattened flow \label{tail}}
\begin{figure}[b]
  \includegraphics[width=0.4\textwidth]{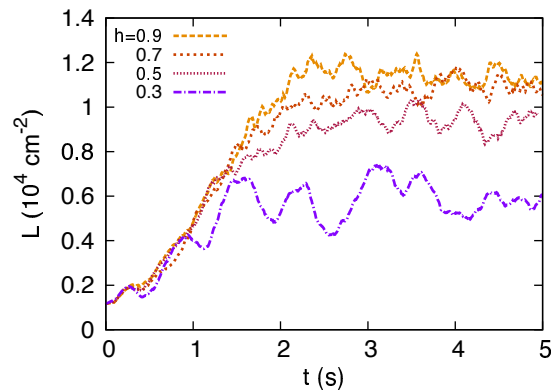}
  \caption{
    (Color online)
    Vortex line density under  tail-flattened flow with various values of $h.$
    The temperature is $T=1.9 ~\mathrm{K}$, and the normal fluid velocity is $v_n=0.5 ~\mathrm{cm/s}$.
  }
  \label{fla_vld_all}
\end{figure}

\begin{figure}[ht]
  \includegraphics[width=0.45\textwidth]{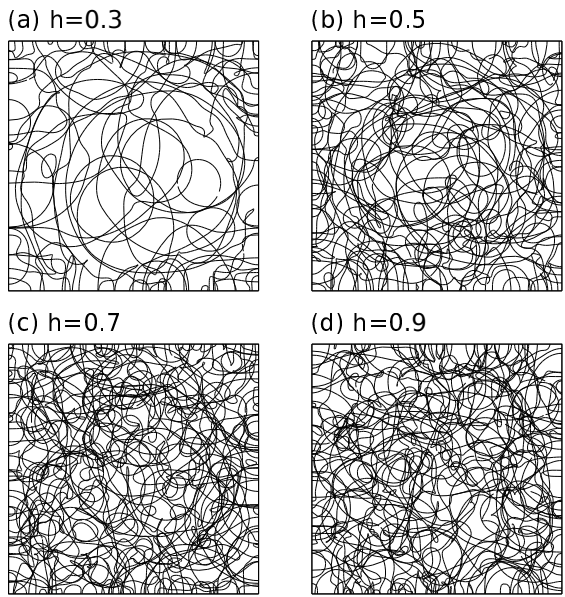}
  \caption{
    Snapshots of the vortex tangle in  tail-flattened flow viewed along the flow direction with various values of $h$.
    These correspond to $t=4.00 ~\mathrm{s}$ of Fig. \ref{fla_vld_all}
  }
  \label{tangles_tail}
\end{figure}

Our model of the tail-flattened flow of Eq. (\ref{fla}) enables us to study how the vortex tangle is affected when the flow profile of the normal fluid is changed by controlling $h$.
The value of $h=0$ gives the Hagen-Poiseuille flow discussed in Sec. \ref{poiseuille}.
The flow profile observed by Marakov {\it et al.} \cite{marakov} looks to be represented approximately by our model with $h=0.7$.
The uniform flow profile studied  in most numerical works corresponds to $h=1$.
Figure \ref{fla_vld_all} shows how $L$ grows for different values of $h$ for $v_{n}=0.5 ~\mathrm{cm/s}$ and $T=1.9 ~\mathrm{K}$.
The large-amplitude oscillation   still remains for low values of $h$, whereas it disappears as the flow profile becomes more uniform with increasing $h$.
It is interesting to find that the statistically steady values of $L$ are almost saturated for $h \gtrsim 0.7,$ although the flow profile is not yet uniform.
Figures \ref{tangles_tail} shows snapshots of the vortex tangles for four different values of $h$.
With increasing $h$, the vortex tangle becomes dense and homogeneous.
Highly curved vortices appear at higher $h$, meaning that reconnection occurs frequently.
Visually, the vortex tangle is almost homogeneous above $h=0.7$, and the two configurations of $h=0.7$ and $h=0.9$ look similar.
We confine ourselves in this section to the case of $h=0.7$ and $T=1.9 ~\mathrm{K,}$ because in this case the flow profile mimics that observed by Marakov {\it et al.} and should give a typical transitional state between  Hagen-Poiseuille flow and  uniform flow.

First, we investigate the statistics of the quantum turbulence in statistically steady states.
Here every physical quantity is averaged temporally over the statistically steady states and spatially over the computational box.
One finds that they are similar to those of uniform counterflow.

\begin{figure}[b]
  \includegraphics[width=0.4\textwidth]{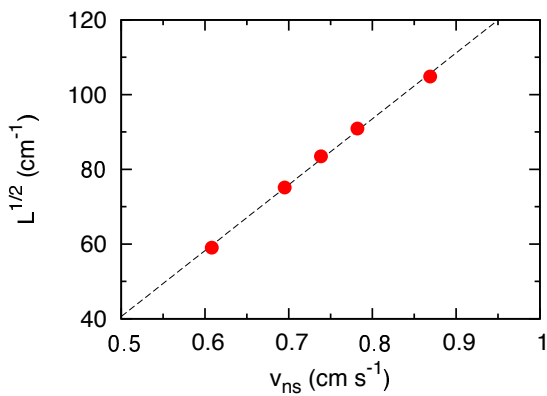}
  \caption
  {
    (Color online)
    Vortex line density averaged over the statistically steady state as a function of $v_{ns}$ under tail-flattened flow.
    The value of $\gamma_{\rm tf}$ is $176 ~\mathrm{s~cm^{-2}}$ for $h=0.7$ and $T=1.9 ~\mathrm{K}$.
    The dashed line refers to the fitting of the data.
  }
  \label{fla_vld_ave.eps}
\end{figure}

Figure \ref{fla_vld_ave.eps} shows $L$, which almost satisfies the relation of Eq. (\ref{vinen}).
We regard the counterflow velocity $v_{ns}$ as the spatially averaged amplitude of ${\bm v}_{ns}={\bm v}_{n}-{\bm v}_{s,a}$.
The value of $\gamma_{\rm tf}$ obtained from this simulation is $176 ~\mathrm{s~cm^{-2}}$, which is larger than those of $\gamma_{\rm hp}=103 ~\mathrm{s~cm^{-2}}$ and $\gamma_{\rm uni} =140.1~\mathrm{s~cm^{-2}}$.
Because the tail-flattened profile with $h=0.7$ is almost uniform, the origin of the difference in $\gamma$ might just be the solid boundary.
This suggests that the solid boundary increases $\gamma$.
However, it is unknown how the solid boundary increases the value of $\gamma$.
The parameter $v_0$ is $\sim 0.3 ~\mathrm{cm/s}$, which is between that of  Hagen-Poiseuille flow and that of  uniform flow \cite{adachi}.
This is consistent with the fact that tail-flattened flow is their intermediate state.

\begin{figure}[ht]
  \includegraphics[width=0.4\textwidth]{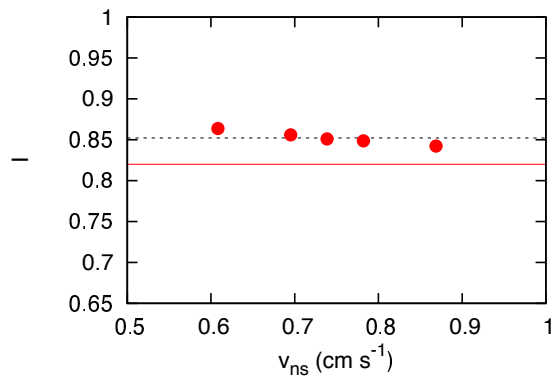}
  \caption
  {
    (Color online)
    Anisotropic parameter $I$ averaged over the statistically steady state under  tail-flattened flow as a function of $v_{ns}$.
    The dotted line are the value of $I$ averaged over the $v_{ns}$.
    The red solid line is the mean value of $I$ obtained by Adachi {\it et al.} \cite{adachi} with $T=1.9 ~\mathrm{K}$.
  }
  \label{fla_iso_ave.eps}
\end{figure}

Figure \ref{fla_iso_ave.eps} shows the anisotropic parameter $I$ versus the counterflow velocity $v_{ns}$.
The anisotropy is almost independent of $v_{ns}$, in agreement with experimental observations \cite{wang}.
The dotted line shows the values of $I$ averaged over $v_{ns}$.
The red solid line shows the mean values of $I$ obtained by Adachi \cite{adachi} {\it et al.}
These results show slightly higher values of $I$ than those obtained from the simulation subject to  uniform counterflow \cite{adachi}.
This is because  the Hagen-Poiseuille profile around the central region increases the anisotropy.

Second, we analyze the local physical quantities.
Every physical quantity is averaged temporally over the statistically steady states and spatially over the flow direction.
The characteristic inhomogeneous behavior appears, whereas the physical quantities averaged over the whole volume are similar to those of uniform counterflow.

\begin{figure}[b]
  \includegraphics[width=0.4\textwidth]{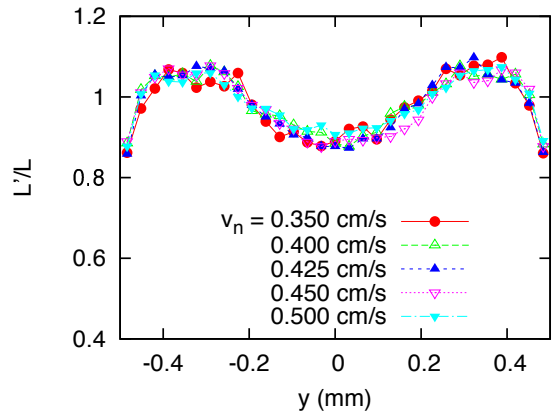}
  \caption
  {
    (Color online)
    Profile of the vortex line density $L'$ normalized by $L$ along $z=0 ~\mathrm{mm}$ under tail-flattened flow.
  }
  \label{fla_vld3.eps}
\end{figure}

Figure \ref{fla_vld3.eps} shows the spatially dependence of $L'/L$.
We plot the profiles for five different values of $v_n$, and the profile is almost independent of $v_n$.
This shows the distribution of the vortices along $z=0 ~\mathrm{mm}$.
 $L'/L$ decreases in the central region and very near the channel walls.
The decrease in the central region is caused by the remaining Hagen-Poiseuille profile, as discussed in Sec. \ref{poiseuille}.
The value of $L'$ increases in the flat region $|y| >0.25$ of $u_t$.
At the channel wall, $L'$ decreases, as discussed in Sec. \ref{poiseuille}.
The vortex-concentrated region near the channel walls is a boundary layer in the tail-flattened flow.
The boundary layer is broader and gentler than in the case of  Hagen-Poiseuille flow.

\begin{figure}[t]
  \includegraphics[width=0.4\textwidth]{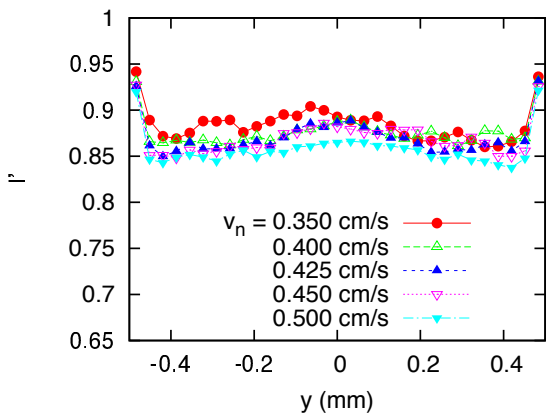}
  \caption
  {
    (Color online)
    Profiles of the anisotropic parameter $I'$ along $z=0 ~\mathrm{mm}$ under tail-flattened flow.
  }
  \label{fla_iso3.eps}
\end{figure}

Figure \ref{fla_iso3.eps} shows the spatial dependence of the local anisotropic parameter $I'$ along $z=0 ~\mathrm{mm}$.
We plot the profiles for five different values of $v_n$, and the profile is almost independent of $v_n$.
The profile is almost uniform, but $I'$ increases slightly in the central region, where the remaining Hagen-Poiseuille profile increases anisotropy.
At the channel walls, $I'$ increases significantly.
This comes from the solid boundary condition for the superfluid.

\begin{figure}[b]
  \includegraphics[width=0.4\textwidth]{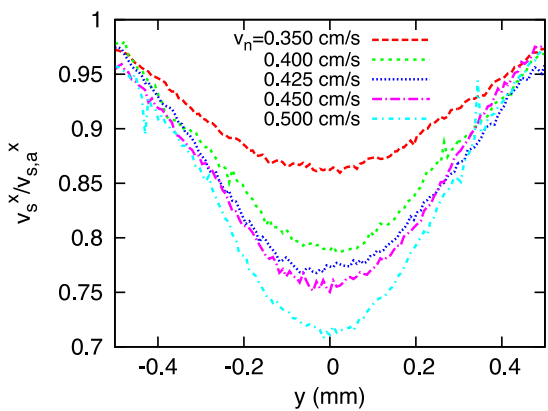}
  \caption
  {
    (Color online)
    Flow direction component $v_x^s$ of the superfluid velocity filed ${\bm v}_{s}$ along $z=0 ~\mathrm{mm}$ under tail-flattened flow.
    These values are normalized by $v_{s,a}^x$.
  }
  \label{fla_prt2.eps}
\end{figure}

Finally, we study the superfluid velocity field.
The flow direction component $v_s^x$ of the superfluid velocity field ${\bm v}_s$ is plotted along $z=0 ~\mathrm{mm}$ in Fig. \ref{fla_prt2.eps}.
These values are normalized by $v_{s,a}^x$.
The values are averaged temporally over the statistically steady states and spatially over the flow direction.
We plot data for $T=1.9 ~\mathrm{K}$ and five different velocities.
The superfluid flow is reduced in the center of the channel.
A velocity reduction is similarly observed in  Hagen-Poiseuille flow, although the two profiles are different.
The velocity reduction in the central region becomes greater with increasing $v_n$.
This comes from the fact that $L$ increases with $v_n$, so that the vorticity in the boundary layer increases with $v_n$.
In contrast, the velocity in the outer region $|y| > 0.25$ is insensitive to $v_n$.
This might be a property of the velocity in the boundary layer.

%**************************************************************************************************
%**************************************************************************************************
\section{Discussion\label{discussions}}
We discuss five important topics by using the results of this work.

%--------------------------------------------------------------------------------------------------
\subsection{Effects of the square channel geometry \label{effects}}
It is important to understand how the geometry of the square channel affects quantum turbulence.
Of course, a real channel is not the periodic cube used in the preceding studies \cite{adachi}.
Some phenomena, such as the T1-T2 transition, depend on the channel geometry.
Because we have focused on the profiles of the physical quantities along $z=0 ~\mathrm{mm}$, properties attributed to the square channel geometry have been missed.
We investigate the profiles of the physical quantities over the channel cross section to consider the geometry effect.

\begin{figure}[ht]
  \includegraphics[width=0.4\textwidth]{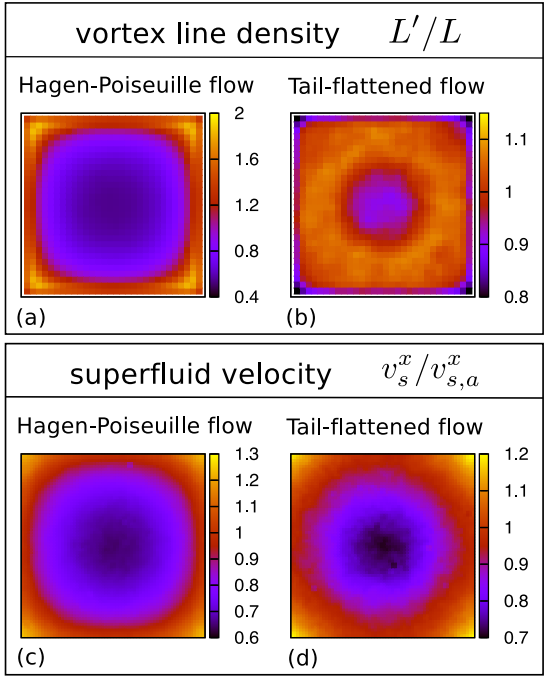}
  \caption{
    (Color online)
    Profiles of vortex line density $L'$ and the flow direction component $v_s^x$ of the superfluid velocity ${\bm v}_s$ over the channel cross section.
    The temperature is $1.9 ~\mathrm{K}$ for all figures.
    The normal fluid velocity is $v_n =0.75 ~\mathrm{cm/s}$ for  Hagen-Poiseuille flow and $v_n=0.5 ~\mathrm{cm/s}$ for  tail-flattened flow.
  }
  \label{2d_profiles}
\end{figure}

Figure \ref{2d_profiles} shows profiles of the local vortex line density and the superfluid velocity over the channel cross section.
The quantities are averaged over the statistically steady states.
We compare the typical results of  Hagen-Poiseuille flow and  tail-flattened flow.
Figures \ref{2d_profiles}(a) and \ref{2d_profiles}(b) show the profiles of the local vortex line density $L'/L$ averaged over the flow direction.
Under  Hagen-Poiseuille flow the vortices concentrate strongly in the channel corners.
This is a consequence of the quantized vortices tending to remain in the corners, because mutual friction is weaker in the region and the vortices neither like to expand nor shrink in the corners.
In contrast, under tail-flattened flow the vortex line density is almost uniform in the flat region.
In this system, vortices cease to remain in the corners, because  mutual friction is almost uniform over that region.
Figures \ref{2d_profiles}(c) and \ref{2d_profiles}(d) show profiles of the flow direction component $v_s^x$ of the superfluid velocity ${\bm v}_s$.
Under  Hagen-Poiseuille flow the value of $v_s^x$ becomes a minimum in the center of the channel, and increases with approaching the channel walls.
In the corners, the superfluid velocity becomes larger than the applied one.
Under tail-flattened flow the velocity reduction becomes greater in the central region than in the case of  Hagen-Poiseuille flow.

%--------------------------------------------------------------------------------------------------
\subsection{Velocity of vortices}
\begin{figure}[ht]
  \includegraphics[width=0.45\textwidth]{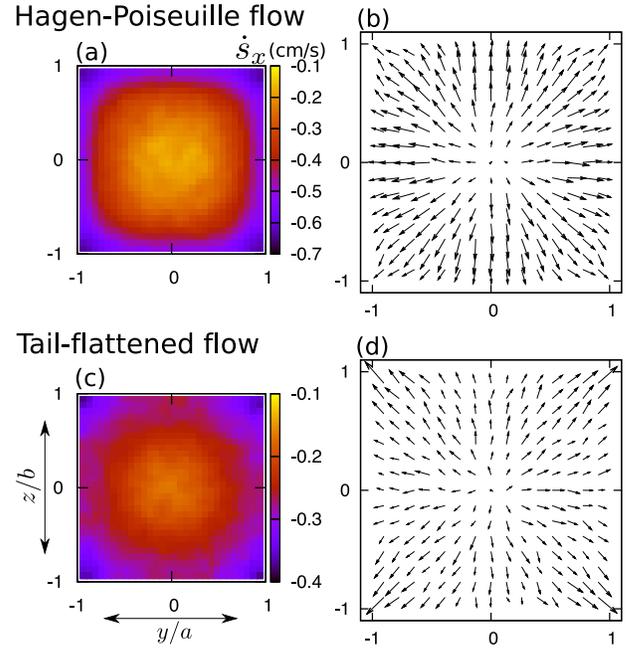}
  \caption{
    (Color online)
    Profiles of velocity $\dot{\bm{s}}$ of quantized vortices.
    The left figures show the component $\dot{s}_x$ along the  flow direction.
    The right figures plot the vector field $(\dot{s}_y,\dot{s}_z)$.
    The temperature is $1.9 ~\mathrm{K}$ for all figures.
    The normal fluid velocity is $v_n =0.75 ~\mathrm{cm/s}$ for  Hagen-Poiseuille flow and $v_n=0.5 ~\mathrm{cm/s}$ for  tail-flattened flow.
    The values of $(\dot{s}_y^2+\dot{s}_z^2)^{1/2}$ averaged over the whole volume are about $0.13 ~\mathrm{cm/s}$ for (b) and $0.05 ~\mathrm{cm/s}$ for (d).
  }
  \label{drift}
\end{figure}

It is also important to know how the vortices move under the nonuniform counterflow.
Thus we calculated the velocity $\dot{\bm{s}}$ of the vortices [Eq. (\ref{vortex})] over the channel cross section.
Figures \ref{drift} (a) and (b) show $\dot{s}_x$ and $(\dot{s}_y, \dot{s}_z)$ of Hagen-Poiseuille flow.
These data was obtained by the average of the statistically steady states and the average over the flow direction.
The average of $(\dot{s}_y^2+\dot{s}_z^2)^{1/2}$ over the cross section is about $0.13 ~\mathrm{cm/s}$.
This figure reveals the characteristic motion of vortices.
Since $\dot{s}_x$ is negative everywhere in Fig. \ref{drift} (a), the vortices turn out to move basically along $\bm{v}_{s,a}$. 
Since $v_{s,a} =-(\rho_n/\rho_s)v_n =-0.57 ~\mathrm{cm/s}$, the vortices move slower than $v_{s,a}$ in the center but faster near the channel walls.
As shown in Fig. \ref{drift} (b), the vortices like to move from the center towards the walls.
The similar behavior appears in Fig. \ref{drift} (c) and (d) of tail-flattened flow too.
However, the velocity amplitude is generally smaller than that of Hagen-Poiseuille flow.

%--------------------------------------------------------------------------------------------------
\subsection{Modification of counterflow velocity \label{modification}}
In Secs. \ref{poiseuille} and \ref{tail} we regarded the counterflow velocity as ${\bm v}_{ns}={\bm v}_n - {\bm v}_{s,a}$ in the analysis of $\gamma$.
However,  the actual relative velocity obtained from Eq. (\ref{mass}) is ${\bm v}_{ns} = {\bm v}_{n} - {\bm v}_{s}$, and, in  nonuniform counterflow, $\overline{v_s} \ne v_{s,a}$, as shown in those sections.
Thus we should recalculate the counterflow velocity by taking account of the reduction in the velocity of the superfluid.

\begin{figure}[b]
  \includegraphics[width=0.45\textwidth]{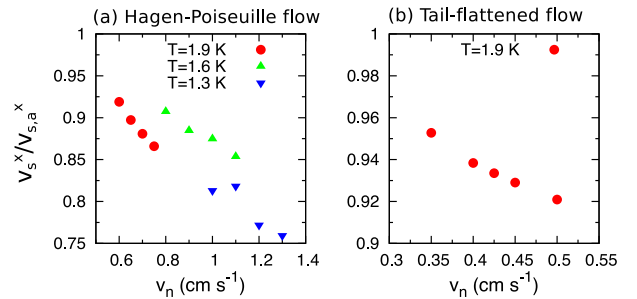}
  \caption
  {
    (Color online)
    Flow direction component $v_s^x$ of the superfluid velocity field ${\bm v}_s$ averaged over the channel cross section under (a)  Hagen-Poiseuille flow and (b)  tail-flattened flow.
    The value of $v_s^x$ decreases with $v_n$, meaning that breakdown of mass conservation becomes larger.
  }
  \label{vs.eps}
\end{figure}

Figure \ref{vs.eps} shows the flow direction component $v_s^x$ of the superfluid velocity ${\bm v}_s$ averaged over the channel cross section and over the statistically steady state.
Figure \ref{vs.eps}(a) show the case under  Hagen-Poiseuille flow, and Fig. \ref{vs.eps}(b) shows that under tail-flattened flow.
In Fig. \ref{vs.eps}(a), the values of $v_s/v_{s,a}$ for $T=1.6 ~\mathrm{K}$ tend to be larger than those for $T=1.3 $ and $1.9 ~\mathrm{K}$.
The values generally decrease from unity, and this means that  mass conservation [Eq. (\ref{mass})] is broken.
The breakdown becomes larger as $v_n$ increases, because $v_s/v_{s,a}$ decreases with $v_n$.
This breakdown shows that the formulation of this work is not sufficient for  correctly describing  thermal counterflow in a realistic channel.
Thus some procedure is required to  dynamically modify the velocities of the two fluids to satisfy  mass conservation.

\begin{table}[t]
  \caption{
    Modified line density coefficients $\gamma$.
    Given are numerical results with  Hagen-Poiseuille flow, $\gamma^*_{\rm hp}$, and with  tail-flattened flow, $\gamma^*_{\rm tf}$,  obtained in this work.
  }
  \begin{tabular}{ccc}
    \hline \hline
    $T$& $\gamma^*_{\rm hp}$ & $\gamma^*_{\rm tf}$  \\
    (K)& (s/cm$^2$) &(s/cm$^2$)  \\
    \hline
      1.3 & 32     & ---    \\
      1.6 & 50     & ---    \\
      1.9 & 123    & 189    \\
    \hline \hline
  \end{tabular}
  \label{gamma2}
\end{table}

By using these results we calculate $v_{ns} = \overline{|{\bm v}_n - {\bm v}_s|}$.
Table \ref{gamma2} lists the values of modified $\gamma$  for Hagen-Poiseuille flow, $\gamma^*_{\rm hp}$, and for tail-flattened flow, $\gamma^*_{\rm tf}$.
The values of $\gamma^*$ are larger than the values of $\gamma$, because the reduction in $v_s$ becomes greater with increasing $v_n$.
The difference between $\gamma$ and $\gamma^*$ becomes smaller as temperature decreases.
This is because  $\overline{v_n}/\overline{v_s} = \rho_s /\rho_n$ increases with $T$, so that the contribution from the reduction in $v_s$ becomes smaller.

%--------------------------------------------------------------------------------------------------
\subsection{Aspect ratio of channel cross section \label{aspect}}
The present work addresses only the case in which the aspect ratio of the channel is unity, but the change of the ratio should seriously affect the profile of the vortex line density.
The aspect ratio of the channel plays an important role in the density and the profile of the vortex tangle.
As long as the counterflow velocity is fixed, the increase of the aspect ratio from unity reduces the counterflow velocity gradient along the long side of the cross section.
Thus the exclusion of the vortex tangle shown in Figs. \ref{tangles}(d) does not occur so much along the long side of the cross section.
Hence the vortex line density should increase compared to the vortex line density in a low-aspect-ratio channel.

It would be meaningful to consider how the aspect ratio affects the  normal fluid, though it is prescribed in this formulation.
A linear stability analysis of the Navier-Stokes equation shows that the critical Reynolds number for turbulence transition of a viscous fluid increases significantly with decreasing  aspect ratio \cite{tatsumi}.
We expect that  the normal fluid in counterflow could remain laminar in a low-aspect-ratio channel even if the superfluid becomes turbulent and the vortex tangle disturbs the normal fluid, which may correspond to the T1 state.
To understand the T1 state, therefore,  studies in a low-aspect-ratio channel  will be indispensable.

%--------------------------------------------------------------------------------------------------
\subsection{T1-T2 transition \label{t1}}
Considering the report \cite{marakov} by Marakov {\it et al.} we would expect  the T1-T2 transition to be a transition of the normal fluid from  tail-flattened flow to  turbulent flow.
Here we discuss the T1-T2 transition under this assumption.
The nature of the T1 state can be understood by using the results of tail-flattened flow obtained in this work.
The T1 state will be the about homogeneous quantum turbulence.
Under  tail-flattened flow, the vortex line density increases with flat parameter $h$, becoming saturated at $h \gtrsim 0.7$.
This suggests that the vortex line density does not jump at the transition from tail-flattened flow to turbulent flow.
According to experiments \cite{tough}, at the T1-T2 transition the vortex line density does not jump, in agreement with our results.
The value of $\gamma_{\rm tf}$ obtained in this work is larger than $\gamma$ of the T1 state obtained in the experiment.
The origin of this difference is unknown, but  addressing this issue by using coupled equations for the two fluids would resolve this discrepancy.

%**************************************************************************************************
%**************************************************************************************************
\section{Conclusions\label{conclusions}}
In this study, we have investigated counterflow quantum turbulence with  nonuniform flows of the normal fluid using the vortex filament model.
The velocity field of the normal fluid was prescribed to consist of two nonuniform profiles, and the solid boundary condition is applied to the walls of a square channel.

First, we conducted a simulation with Hagen-Poiseuille flow.
The vortex line density has a fluctuation with a large amplitude.
We investigated the configuration of the vortex tangle in the oscillation, and it was revealed that one cycle of the oscillation consists of four characteristic  states.
The relation $L^{\frac{1}{2}} = \gamma (v_{ns} - v_0)$ is almost satisfied, but there are quantitative discrepancies between the results of this work and those found in uniform counterflow.
The anisotropy found here is larger than in the case of uniform counterflow.
Then we investigated the inhomogeneity of the vortex tangle.
The quantized vortices concentrate near the channel walls, and the vortex tangle becomes more isotropic near the solid boundaries.
The region near the solid boundary is a boundary layer of the superfluid, which was also found by Baggaley {\it et al.} \cite{baggaley13}$^{,}$\cite{baggaley14}.
We investigated the superfluid velocity field.
It is found that a velocity field ${\bm v}_{s,\omega}$ opposite to the applied flow ${\bm v}_{s,a}$ is created by the inhomogeneous vortex tangle.
Thus the superfluid flow becomes smaller than the applied one.

Second, we conducted a simulation with tail-flattened flow.
We showed that the vortex line density increases significantly as the flat parameter $h$ increases, becoming saturated at $h \gtrsim 0.7$.
We confined ourselves to $h=0.7$ to be consistent with the experiments.
The value of $\gamma$ was found to be larger than those obtained from simulations with  uniform counterflow, and
the anisotropy was found to be slightly larger than in the case of  uniform counterflow.
The inhomogeneity is significantly reduced  in comparison to the case of  Hagen-Poiseuille flow.
In the center of the channel the vortex line density becomes slightly lower, and the vortex tangle becomes slightly anisotropic.
This is because the Hagen-Poiseuille profile remains in the central region of the tail-flattened profile.
The velocity field ${\bm v}_{s,\omega}$ is different from that in the case of Hagen-Poiseuille flow, where the reduction of the central region becomes greater.

The present simulation revealed that tail-flattened flow is some intermediate state between  Hagen-Poiseuille flow and  uniform flow.
When we focus on the physics quantities averaged over the whole volume, we find that they are similar to those of uniform counterflow.
However, if we look at the spatial dependence of these values, they show  inhomogeneous behavior.
These considerations suggest that some {\it in situ} observation is necessary  to understand tail-flattened flow.

%**************************************************************************************************
%**************************************************************************************************
\acknowledgments
We would like to acknowledge W. F. Vinen and Wei Guo for useful discussions.
This work was supported by JSPS KAKENHI Grant No. 26400366 and MEXT KAKENHI ``Fluctuation \& Structure" Grant No. 26103526.

%**************************************************************************************************
%**************************************************************************************************

%**************************************************************************************************
%**************************************************************************************************

\begin{thebibliography}{99}
  \bibitem{tough}J. T. Tough, in {\it Progress in Low Temperature Physics}, edited by D. F. Brewer (North-Holland, Amsterdam, 1982), Vol. VIII.
  \bibitem{schwarz88} K. W. Schwarz, Phys. Rev. B {\bf 38}, 2398 (1988).
  \bibitem{adachi} H. Adachi, S. Fujiyama, and M. Tsubota, Phys. Rev. B {\bf 81}, 104511 (2010).
  \bibitem{marakov} A. Marakov, J. Gao, W. Guo, S. W. Van Sciver, G. G. Ihas, D. N. McKinsey, and W. F. Vinen, Phys. Rev. B {\bf 91}, 094503 (2015).
  \bibitem{guo09} W. Guo, J. D. Wright, S. B. Cahn, J. A. Nikkel, and D. N. McKinsey, Phys. Rev. Lett. {\bf 102}, 235301 (2009).
  \bibitem{guo10} W. Guo, S. B. Cahn, J. A. Nikkel, W. F. Vinen, and D. N. McKinsey, Phys. Rev. Lett. {\bf 105}, 045301 (2010).
  \bibitem{donnelly} R. J. Donnelly, in {\it Quantized Vortices in Helium {\rm II}}, edited by A. M. Goldman {\it et al.} (Cambridge University Press, Cambridge, England, 1991).
  \bibitem{onsager} L. Onsager, Nuovo Cimento Suppl. {\bf 6}, 279 (1949).
  \bibitem{vinen61} W. F. Vinen, Proc. R. Soc. London Ser. A {\bf 260}, 218 (1961).
  \bibitem{vinen57} W. F. Vinen, Proc. R. Soc. London Ser. A {\bf 242}, 493 (1957).
  \bibitem{melotte} D. J. Melotte and C. F. Barenghi, Phys. Rev. Lett. {\bf 80}, 4181 (1998).
  \bibitem{bewley} G. P. Bewley, D. P. Lathrop, and K. R. Sreenivasan, Nature (London) \textbf{441}, 588 (2006).
  \bibitem{paoletti} M. S. Paoletti, R. B. Fiorito, K, R. Sreenivasan, and D. P. Lathrop, J. Phys. Soc. Jpn. {\bf 77}, 111007 (2008).
  \bibitem{zhang} T. Zhang and S. W. Van Sciver, Nat. Phys. {\bf 1}, 36 (2005).
  \bibitem{mantia} M. La Mantia and L. Skrbek, Phys. Rev. B {\bf 90}, 014519 (2014).
  \bibitem{samuels} D. C. Samuels, Phys. Rev. B {\bf 46}, 11714 (1992).
  \bibitem{aarts} R. G. K. M. Aarts and A. T. A. M. de Waele, Phys. Rev. B {\bf 50}, 10069 (1994).
  \bibitem{baggaley13} A. W. Baggaley and S. Laizet, Phys. Fluids {\bf 25}, 115101 (2013).
  \bibitem{baggaley14} A. W. Baggaley and S. Laurie, J. Low Temp. Phys. {\bf 178}, 35 (2014).
  \bibitem{kivotides} D. Kivotides, Phys. Rev. B {\bf 76}, 054503 (2007).
  \bibitem{yui} S. Yui and M. Tsubota, J. Phys.: Conf. Ser. {\bf 568}, 012028 (2014).
  \bibitem{schwarz85} K. W. Schwarz, Phys. Rev. B {\bf 31}, 5782 (1985).
  \bibitem{poiseuille} {\it The handbook of fluid dynamics}, edited by R. W. Johnson (CRC Press, Boca Raton, 1998).
  \bibitem{barenghi} C. F. Barenghi, R. J. Donnelly, and W. F. Vinen, J. Low Temp. Phys. {\bf 52}, 189 (1983).
  \bibitem{childers} R. K. Childers and J. T. Tough, Phys. Rev. B {\bf 13}, 1040 (1976).
  \bibitem{wang} R. T. Wang, C. E. Swanson, and R. J. Donnelly, Phys. Rev. B {\bf 36}, 5240 (1987).
  \bibitem{maurer} J. Maurer and P. Tabeling, Europhys. Lett. {\bf 43}, 29 (1998).
  \bibitem{stalp} S. R. Stalp, L. Skrbek, and R. J. Donnelly, Phys. Rev. Lett. {\bf 82}, 4831 (1999).
  \bibitem{tatsumi} T. Tatsumi and T. Yoshimura, J. Fluid Mech. {\bf 212}, 437 (1990).
\end{thebibliography}
\end{document}